\documentclass[aps,prl,twocolumn,groupedaddress,showpacs,endfloat*,secnumarabic,amssymb,amsmath,tightenlines]{revtex4}

\usepackage{graphicx}

\begin{document}
\title{A New Type of Two-photon Forward Radiation in Pure Liquids}
\author{Hong-tao Bian}
\thanks{Also Graduate School of the Chinese Academy of Sciences}
\author{Yi Rao}
\thanks{Current address: Department of Chemistry, Columbia University, New York, NY 10025, U.S.A.}
\author{Yan-yan Xu}
\thanks{Also Graduate School of the Chinese Academy of Sciences}
\author{An-an Liu}
\thanks{Also Graduate School of the Chinese Academy of Sciences}
\author{Yuan Guo}
\author{Hong-fei Wang}
\email{hongfei@iccas.ac.cn}
\thanks{Author to whom correspondence
should be addressed.} \affiliation{State Key Laboratory of
Molecular Reaction Dynamics,
\\Institute of Chemistry, the Chinese Academy of Sciences,
Beijing, China, 100080}

\date{\today}

\begin{abstract}

Unexpected spectral features are observed in the two photon spectrum
of the pure water in the forward direction when an 80 femtosecond
laser pulse is focused at $10^{10}Wcm^{-2}$ or less. Such intensity
is much lower than the breakdown or stimulated threshold of the
liquid water. The two broad features are about $2700cm^{-1}$ and
$5000cm^{-1}$ red shifted from the hyper-Rayleigh wavelength,
respectively, and they are quadratic with the laser intensity. They
do not match the known Raman or hyper-Raman frequencies of water,
and they are both centered at a narrow angle in the forward
direction. Several other liquids also exhibited similar but
molecular specific spectral features.

\end{abstract}


\maketitle

Different optical processes can be observed when intense laser
pulses with different intensities are propagated in the condensed or
gaseous media. Incoherent processes such as Rayleigh and Raman
scattering can be observed in all directions, while most of the
coherent or stimulated processes, such as stimulated Raman,
dielectric breakdown, self focusing and continuum generation, are
generally centered in the forward direction.\cite{BoydBook,Shenbook}

Here we report the observation of the unexpected spectral features
in the two photon spectrum of the pure water in the forward
direction when an 80 femtosecond laser pulse is focused into liquid
water at the intensity of $10^{10}Wcm^{-2}$ or less. The two
observed broad features are about $2700cm^{-1}$ and $5000cm^{-1}$
red shifted from the hyper-Rayleigh wavelength, respectively, and
they are quadratically dependent on the input laser intensity. To
our surprise, they do not match the known Raman or hyper-Raman
frequencies of water, and they are both centered at a narrow angle
in the forward direction. Several other liquids also exhibited
similar but molecular specific spectral features. The difference of
the spectral shift between the water and the heavy water might be
attributed to the isotope effect. Since the laser intensity here is
much lower than the breakdown or stimulated threshold of the liquid
water,\cite{BloembergenPRA,LiuOptCommun,SawadaPRL4110} the mechanism
of this new type of two photon forward radiation is yet to be
determined.

\begin{figure}[h!]
\begin{center}
\includegraphics[height=10cm,width=6.5cm]{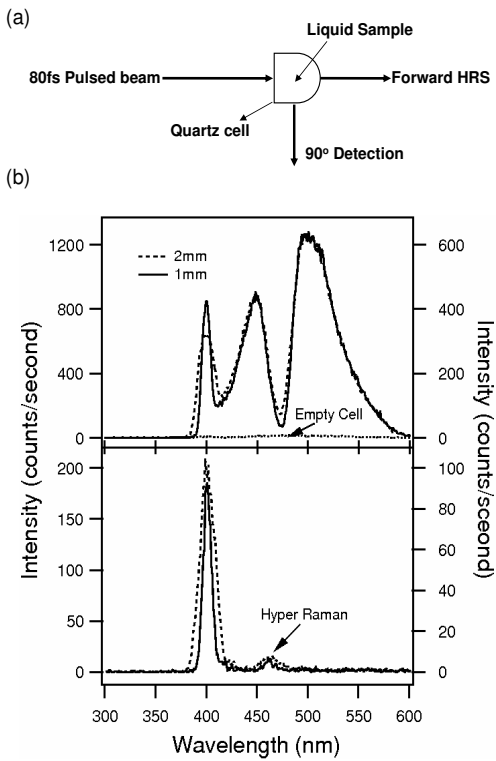}
\caption{(a) Schematic illustration of the forward direction and 90
degree direction hyper-rayleigh scattering. (b) Spectra of the pure
water in the forward direction (Top) and in the $90^{o}$ direction
(Bottom) with the fundamental wavelength at 800nm. The laser power
is 500mW. The solid line and dot line denote the monochromator with
1mm and 2mm slit. Both the input and output polarization are p
polarized (vertical).}\label{Water800nmSpectroscopy}
\end{center}
\end{figure}

The schematic of the angular and spectral resolved two-photon
scattering measurement in Fig. \ref{Water800nmSpectroscopy}(a) was
similar to the setup reported previously.\cite{RaoyiJPCA} A
broadband tunable mode-locked Ti:Sapphire laser (Tsunami 3960C) is
pumped by a 10W CW laser (Millennia Xs), producing 80fs pulses at 82
MHz repetition rate. The typical laser power is 500mW at 800nm with
a 9nm bandwidth. A 50mm focal lens focuses the laser beam into a
D-shaped quartz cell with the diameter of 15mm filled with the
sample liquid. The sample cell is fixed at the center position of a
rotating goniometer mounted with the detection optics, enables
detection of the scattered light in different angular directions.
The signal is collected by a condenser lens with $f$=50mm, then
focused into the monochromator (0.1m, Beijing Optical Instrument
Factory WDG10) with a 2nm resolution, and detected with a single
photon counting system (PMT: Hamamatsu PMT R585, Preamplifier:
Stanford Research System SR240, and photon counter: SRS SR400). With
a laser power of 500mW, the power intensity at the beam waist is
$1.0\times10^{10}Wcm^{-2}$. The pure water was prepared from double
distilled water and deionized with the Millipore Simplicity 185
(18.2 M$\Omega$cm). All the experiments were performed at $22^{o}C$.

For the pure water excited at 800nm, the spectral features in the
forward and $90^{o}$ direction are significantly different as in
Fig. \ref{Water800nmSpectroscopy}(b). The signal from the empty cell
is essentially zero. This eliminates the possibility that the
observed signal is generated from any other optical components in
the entire optical path. The hyper-Rayleigh and hyper-Raman peaks in
the $90^{o}$ direction are the same as that in the backward
direction reported by Webb et al. and in the $90^{o}$ direction by
Terhue et al.\cite{WebbAnalChem,TerhunePRLFirst} The hyper-Raman
peak at 463nm corresponds to a Raman shift of $3400cm^{-1}$, which
is vibrational band of the hydrogen bond of the bulk water. The weak
$660cm^{-1}$ and $1600cm^{-1}$ hyper-Raman peaks of the bulk water
reported by Webb et al. are also weak but discernable in our
$90^{o}$ direction spectra.

These are in sharp contrast to the two broad and strong spectral
features at 448nm ($2700cm^{-1}$ red shifted from the sharp 400nm
hyper-Rayleigh band) and 500nm ($5000cm^{-1}$ red shift) in the
forward spectrum. These two peaks correspond to no known Raman or
Hyper-Raman band of the liquid water. Very differently, in other
forward scattering processes, such as stimulated Raman, dielectric
breakdown, laser-induced plasma generation and superbroadening
processes,\cite{SawadaPRL4110,SawadaPRL3512,BloembergenPRA}, the
Stokes Raman band dominated the forward spectra. On the other hand,
the laser intensity in our work is only $10^{10}Wcm^{-2}$ or less,
much less than the thresholds of these other forward scattering
processes.

Denisov et al. discussed the transverse(TO) and longitudinal(LO)
modes of hyper-Raman scattering from the
liquid.\cite{DenisovPhysRep,DenisovOptComm} Basically, both the TO
and the LO modes can be observed in the 90$^o$ direction, while in
the forward direction only the LO modes appear. Therefore, the 448nm
and 500nm bands in the forward direction observed here can not be
attributed to the process similar to the ordinary Raman or
hyper-Raman processes. In addition, they do not fit with the known
Raman and hyper-Raman frequencies. Therefore, we surmise that these
two new bands have to be a new type of forward radiation from the
two-photon excitation.

The polarization of the 400nm, 448nm and 500nm peaks are all
parallel to the input polarization of the fundamental. Using the 1mm
slit instead of the 2mm slit for both the entrance and the exit of
the monochromator, the width of the 400nm hyper-Rayleigh narrows
from 16nm to 9nm, which is the band width of the 800nm femtosecond
pulse. However, the widths of the 448nm and 500nm bands are broad
and remain unchanged. This indicates that the broadening mechanism
for these two new and broad spectral features is quite different
from that of the hyper-rayleigh processes,\cite{SilbeyJCP1561} even
though they follow the same polarization dependence. If the widths
of these two peaks are due to any fast molecular dynamic
interactions, the time scale of these interactions must be in the
range of 30 to 15 femtosecond, respectively.

\begin{figure}[t]
\begin{center}
\includegraphics[height=9cm,width=6cm]{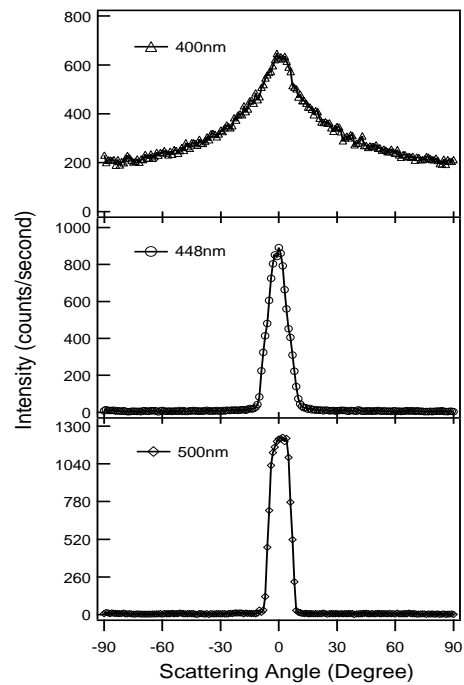}
\caption{Angular resolved scattering angle measurement of the 400nm
(top, open triangle), 448nm (middle, open circle) and 500nm (bottom,
open diamond) peaks in Fig. \ref{Water800nmSpectroscopy}. The slit
width is 2mm.}\label{WaterAngleResloved}
\end{center}
\end{figure}

The angle resolved measurements of the 400nm, 448nm and 500nm peaks
in Fig. \ref{Water800nmSpectroscopy} are shown in Fig.
\ref{WaterAngleResloved}. The angular dependence of the 400nm peak
is again distinctively different from the 448nm and 500nm peaks. As
expected for the hyper-Rayleigh process, the angular dependence here
for the 400nm peaks follows the $\cos^{2}\theta$ function centering
at the forward direction. Therefore, no additional information can
be gained by looking in directions other than the classic 90$^o$
angle for the hyper-rayleigh process.\cite{BersohnJCP,PersoonACR}

On the other hand, the angular dependence of the 448nm and 500nm
peaks is distinctively very narrow and centered at the forward
direction. The two curves may have small difference, but they
certainly imply a very different mechanism from that of the
hyper-Rayleigh or hyper-Raman processes, where no such strong
angular anisotropy has ever been observed, except for processes such
the stimulated Raman or superbroadening.

\begin{figure}[h!]
\begin{center}
\includegraphics[height=4.5cm,width=6cm]{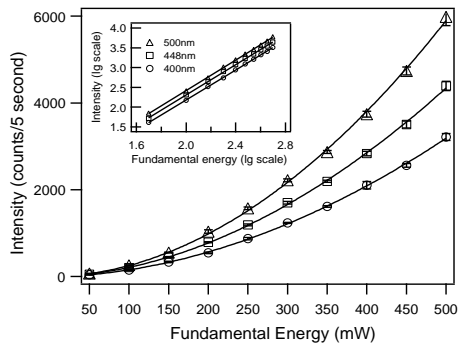}
\caption{Power dependence of the forward 400nm (open circle), 448nm
(open square) and 500nm (open triangle) of the pure water with the
800nm excitation. The solid lines are the fit result using a
quadratic function. The insert shows the corresponding power
dependence with the logarithm scale. }\label{WaterPowerDependence}
\end{center}
\end{figure}

However, the power dependence measurements eliminated the
possibility of any above threshold processes, such as the stimulated
or superbroadening processes. The power dependence of the forward
400nm, 448nm and 500nm bands are all quadratic, as in Fig.
\ref{WaterPowerDependence}. The incident laser power was varied from
50mW to 500mW, i.e. from about $10^{9}Wcm^{-2}$ to
$10^{10}Wcm^{-2}$. The spectral features in Fig.
\ref{Water800nmSpectroscopy} and the angular dependence in Fig.
\ref{WaterAngleResloved} remain unchanged with the laser intensity.
The power index for the three peaks are essentially the same as
$1.9\pm0.1$, indicating a clear two-photon process instead of any
above-threshold behavior.\cite{SchneiderJMS,DasCPL} Furthermore, no
laser-induced breakdown or plasma generation process was
observed.\cite{SawadaPRL4110}

\begin{figure}[h!]
\begin{center}
\includegraphics[height=10cm,width=6.5cm]{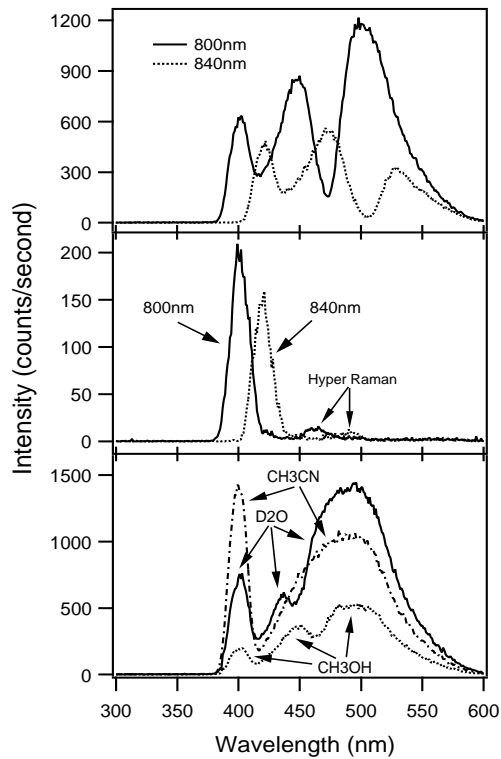}
\caption{Top: Forward spectra of the pure water with the 800nm
(solid line) and the 840nm (dotted line) excitation. Middle:
$90^{o}$ direction spectra of the pure water with the 800nm (solid
line) and the 840nm (dotted line) excitation. Bottom: Forward
spectra of the pure methanol (dotted line), pure heavy water (solid
line) and pure acetonitrile (dashed-dotted line) with the 800nm
excitation.}\label{LiquidForwardSpectroscopy}
\end{center}
\end{figure}

These two new peaks in the forward direction are not from some
unexpected two-photon fluorescence process of the liquid or any
possible impurity, either.\cite{RaoyiJPCA} To demonstrate this, the
excitation fundamental wavelength was changed every 20nm from 740nm
to 880nm, and the positions of these two new peaks shift with the
excitation wavelength just like the ordinary hyper-Rayleigh and
hyper-Raman processes. Data for 800nm and 840nm are shown in the top
and middle panels in Fig. \ref{LiquidForwardSpectroscopy}. However,
the relative intensity of the new peaks to that of the
hyper-Rayleigh peak in the forward direction does change with the
excitation wavelength, as observed for all the liquids measured
below.

Other common liquids also have the similar results as these of the
pure water. The forward spectra of the pure methanol (Sigma-Aldrich,
HPLC grade, 99.9$\%$), heavy water (Acros, NMR grade, 99.95$\%$) and
acetonitrile (Scharlau, HPLC grade, 99.99$\%$) are shown in the
bottom panel of Fig. \ref{LiquidForwardSpectroscopy}. Here the
spectral purity of these liquids are thoroughly examined against
samples from other sources. Similar to water, all the peaks here,
except for the hyper-Rayleigh peak at 400nm, do not match any known
Raman or hyper-Raman frequencies of each molecule. In all these
cases, the intensities of these new spectral features are in the
same order of the forward hyper-Rayleigh scattering intensity.

It is intriguing that these frequencies are molecular or molecular
group specific. The second peak of liquid methanol is at the same
position of the pure water ($2700cm^{-1}$), while the second peak of
the heavy water has a smaller shift ($2100cm^{-1}$) from its
hyper-Rayleigh peak. The ratio between these two frequency shifts is
1.3, in the ballpark of the isotope effect factor for the O-H
vibrational frequencies. However, because both $2700cm^{-1}$ and
$2100cm^{-1}$ correspond to no known vibrational frequencies of
water and heavy water, such value only implies isotope effect, but
may well be just accidental. To test this idea, experiments on pure
ethanol and ethylene glycol were performed. The methanol, ethanol
and ethylene glycol all have the same two peak positions red shifted
from the hyper-Rayleigh peak at 400nm, even though their relative
intensities vary. This seems to indicate that the molecules with O-H
groups have similar peak positions. On the other hand, spectral
features for different molecular groups must be different. To
further support this idea, the new spectral feature found for the
liquid acetonitrile in the same figure is different from all the
above, indicating molecular or molecular group specificity of these
spectral features. Other common liquids, such as $CS_{2}$ and
$CCl_{4}$, were also tested and similar behaviors were observed.
However, the optical purities of the purest samples from the
suppliers have not been as satisfactory as the ones reported here,
and further studies are warranted.

Since there is no such forward spectral features observed in the one
photon Raman studies,\cite{SawadaPRL3512,SawadaPRL4110}, what we
observed is likely a unique process involves at least two photons.
Since the frequencies observed do not match known Raman or
hyper-Raman frequencies, it may come from some collective or
correlated properties of the liquid. Recently, based on some
abnormal depolarization ratios in the hyper-Rayleigh measurement of
the pure liquids in the $90^{o}$ direction, Shelton et al. discussed
the possibility of the long range orientational correlation in pure
liquids.\cite{SheltonJCP849,SheltonPRL,SheltonCPL,SheltonJCP111103}
There is no evidence so far to make the connection between the
abnormal hyper-Rayleigh measurement and the new spectral features
observed here. But certainly they all point to some new
understandings of the molecular and optical physics of the molecular
liquids.

In conclusion, new spectral features from two-photon excitation were
observed in the forward direction from liquid water and other pure
liquids. The intensities of these new spectral features are in the
same order of the forward hyper-Rayleigh scattering intensity. The
laser intensity used in these studies is $10^{10}Wcm^{-2}$ or less,
i.e. well below the threshold for laser-induced breakdown and
stimulated processes in these pure liquids. The strong forward
angular dependence indicates that these features are not from the
simple incoherent linear or nonlinear scattering processes.
Experimental results showed that these new spectral features are
two-photon radiations, but clearly not from the hyper Raman process,
nor the two-photon fluorescence process, and nor the above threshold
processes such as the stimulated processes and the superbroadening
generation. These new spectral features are also unique for
molecules with different molecular structures, and they do not match
any known Raman and hyper-Raman frequencies. To our knowledge, there
is no known mechanism in this range of laser intensity can explain
these experimental observations. Therefore, the origin and the
mechanism of such process warrant further experimental and
theoretical investigations.

HFW is thankful for the support from the National Natural Science
Foundation of China (NSFC No.20425309, No.20533070). YG is thankful
for the support from the National Natural Science Foundation of
China (NSFC No.20425309).


\begin{thebibliography}{99}
\bibitem{BoydBook} R. W. Boyd, Nonlinear Optics, Academic Press, Boston (1992).
\bibitem{Shenbook} Y. R. Shen, The Principles of Nonlinear Optics, Wiley-InterScience, Hoboken (2003).
\bibitem{BloembergenPRA} W. L. Smith, P. Liu and N. Bloembergen,  Phys. Rev. A \textbf{15}, 2396-2403(1977).
\bibitem{LiuOptCommun} W. Liu, S. Petit, A. Becker, N. Akozbek, C. M. Bowden and S. L. Chin,  Opt. Commun. \textbf{202}, 189-197(2002).
\bibitem{SawadaPRL4110} H. Yui, Y. Yoneda, T. Kitamori and T. Sawada,  Phys. Rev. Lett. \textbf{82}, 4110-4113 (1999).
\bibitem{RaoyiJPCA} Y. Rao, X. M. Guo and H. F. Wang,  J. Phys. Chem. A \textbf{108}, 7977-7982 (2004).
\bibitem{TerhunePRLFirst} R. W. Terhune, P. D. Maker and C. M. Savage,  Phys. Rev. Lett. \textbf{14}, 681-684(1965).
\bibitem{WebbAnalChem} C. Xu, J. B. Shear and W. W. Webb,  Anal. Chem. \textbf{69}, 1285-1287 (1997).
\bibitem{SawadaPRL3512} H. Yui and T. Sawada, Phys. Rev. Lett. \textbf{85}, 3512-3515 (2000).
\bibitem{DenisovPhysRep} V. N. Denisov, B. N. Mavrin and V. B. Podobedov,  Phys. Rep. \textbf{151}, 1-92 (1987).
\bibitem{DenisovOptComm} V. N. Denisov, B. N. Mavrin, V. B. Podobedov and Kh. E. Sterin,  Opt. Commun. \textbf{44}, 39-42 (1982).
\bibitem{BersohnJCP} R. Bersohn, Y. H. Pao and H. L. Frisch,  J Chem. Phys. \textbf{45}, 3184-3198 (1966).
\bibitem{PersoonACR} E. Hendrickx, K. Clays and A. Persoons,  Acc. Chem. Res. \textbf{31}, 675-683 (1998).
\bibitem{SchneiderJMS} G. Brehm, G. Sauer, N. Fritz, S. Schneider and S. Zaitsev, J. Mol. Struct. \textbf{735-736}, 85-102 (2005).
\bibitem{DasCPL} K. Das, A. Uppal and P. K. Gupta,  Chem. Phys. Lett. \textbf{426}, 155-158 (2006).
\bibitem{SilbeyJCP1561} S. N. Yaliraki and R. J. Silbey,  J. Chem. Phys. \textbf{111}, 1561-1568 (1999).
\bibitem{SheltonJCP849} P. Kaatz, E. A. Donley and D. P. Shelton,  J. Chem. Phys. \textbf{108}, 849-856 (1998).
\bibitem{SheltonPRL} D. P. Shelton and P. Kaatz,  Phys. Rev. Lett. \textbf{84}, 1224-1227 (2000).
\bibitem{SheltonCPL} D. P. Shelton,  Chem. Phys. Lett. \textbf{325}, 513-516 (2000).
\bibitem{SheltonJCP111103} D. P. Shelton,  J. Chem. Phys. \textbf{123}, 111103 (2005).
\end{thebibliography}
\end{document}